\documentclass[english,pss,pssb]{w-art}
\usepackage{times}
\usepackage[T1]{fontenc}
\usepackage[latin1]{inputenc}
\usepackage{array}
\usepackage{verbatim}
\usepackage{calc}
\usepackage{amsmath}
\usepackage{graphicx}
\usepackage{amssymb}

\makeatletter

\usepackage{cite}

\usepackage{babel}
\makeatother
\begin{document}

\title[
\title{Selective lasing in multimode NPWs}\maketitle
]{Selective lasing in multimode periodic and non-periodic \\ nanopillar waveguides}

\author[S. Zhukovsky]{Sergei V. Zhukovsky\footnote{Corresponding author: e-mail: {\sf sergei@th.physik.uni-bonn.de}, Phone: +49 228 73 5988, Fax: +49 228 73 3223}\inst{1}}

\author[D. Chigrin]{Dmitry N. Chigrin\inst{1}}

\author[A. Lavrinenko]{Andrei V. Lavrinenko\inst{2}}

\author[J. Kroha]{Johann Kroha\inst{1}}

\address{\inst{1} Physikalisches Institut, Universit\"at Bonn, Nussallee
12, D-53115 Bonn, Germany}

\address{\inst{2} COM-DTU, Department of Communications, Optics and Materials,
NanoDTU, Technical University of Denmark, Building 345V, DK-2800 Kgs.
Lyngby, Denmark}



\begin{abstract}
We investigate the lasing action in coupled multi-row 
nanopillar waveguides of periodic or fractal structure using the 
finite difference time domain (FDTD) method, coupled to the laser
rate equations. Such devices exhibit band splitting with distinct
and controllable supermode formation.
We demonstrate that selective lasing into each of the supermodes is
possible. The structure acts as a microlaser with selectable wavelength.
Lasing mode selection is achieved by means of coaxial injection seeding with
a Gaussian signal of appropriate transverse amplitude and phase 
profiles. Based on this we propose the 
concept of switchable lasing as an alternative to conventional
laser tuning by means of external cavity control.
\end{abstract}
\maketitle

\section{Introduction}

Photonic crystals (PhCs) offer a wide range of applications in controlling
the flow of light \cite{Joannopoulos,Sakoda,Loutrioz}. In the photonic
band-gap (PBG) regime, no propagating modes are allowed in
PhCs, making the material ``insulating'' to optical waves. Numerous
applications have been devised on this basis. Creating a line defect
in a PhC structure, e.g., by removing or modifying one or several
lines of scatterers from the PhC periodic lattice, produces a photonic
crystal waveguide (PCW). Various designs of a PCW in two-dimensional
geometry have been reported, based both on air hole lattices
in a dielectric background and on dielectric rods 
in air \cite{JohnsonGuide1,JohnsonGuide2,JapPillar}.
Numerous works have been dedicated to the optimization of PCW-based
integrated optical devices, such as bends, splitters, couplers, etc.
(see, e.g., \cite{Loutrioz}). On the other hand, creating a point-like
defect in a periodic PhC lattice results in a microcavity
supporting exponentially localized defect modes \cite{mcNoda1} with 
a possibly high Q-factor. It
can be used, for instance, as a microresonator for defect-mode lasing
\cite{mcLasing}, or, in combination with the PCW, as add-drop filters
\cite{mcNoda2}.  

\begin{figure}
\begin{tabular}{>{\centering}m{0.4\columnwidth}>{\centering}m{0.6\columnwidth}}
\includegraphics[
clip,width=0.35\columnwidth]{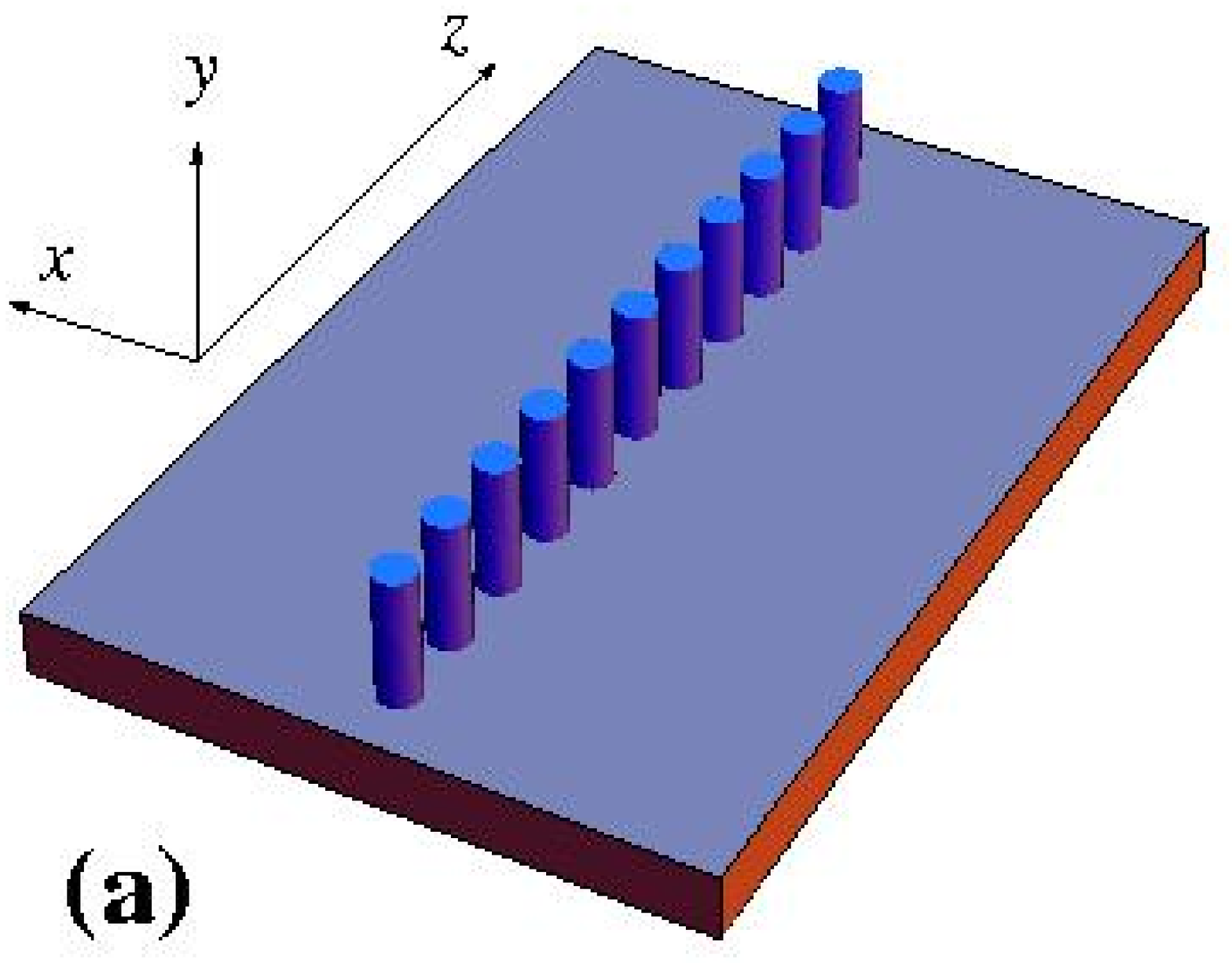}&
\includegraphics[width=0.4\columnwidth]{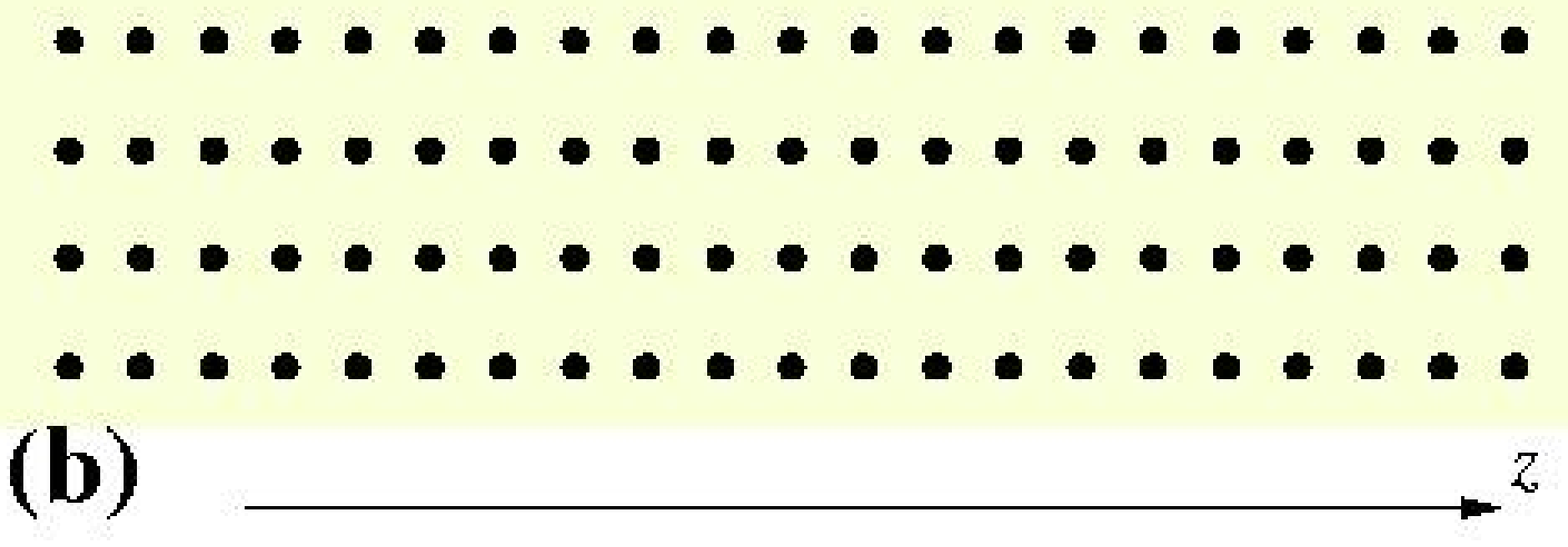} \includegraphics[width=0.4\columnwidth]{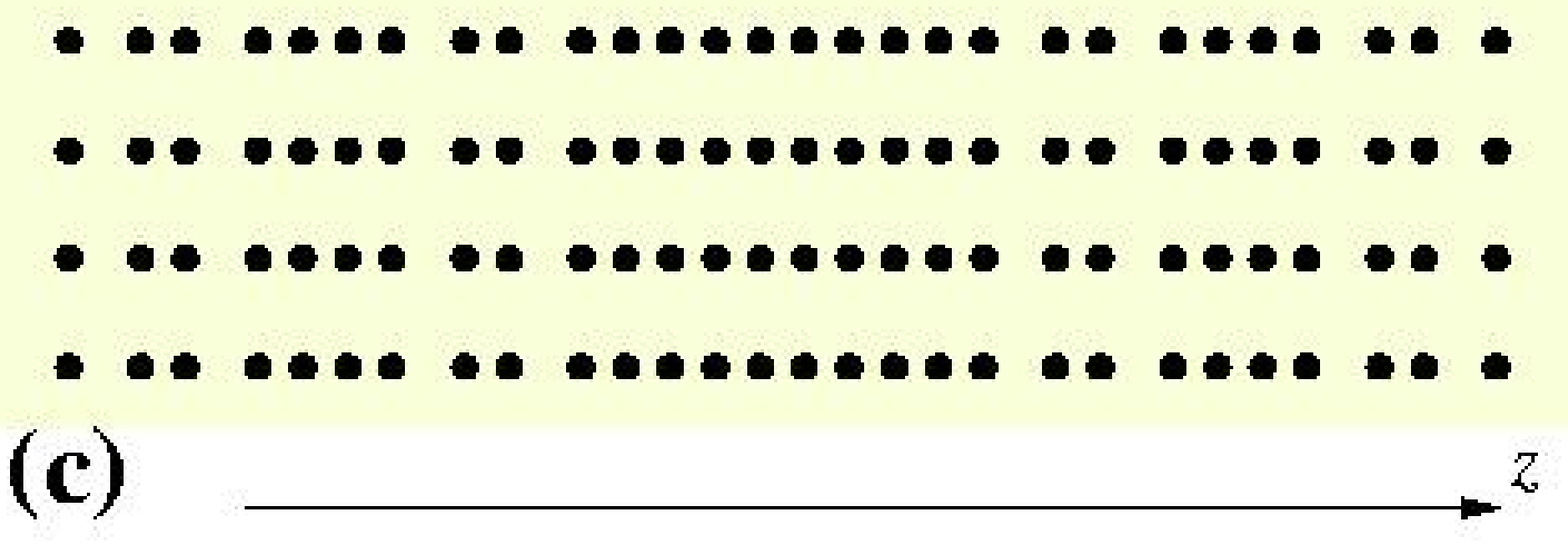}\tabularnewline
\end{tabular}

\caption{\textbf{(a)} Schematic illustration of a periodic nanopillar waveguide
(NPW), along with the top views of the \textbf{(b)} four-row coupled
periodic and \textbf{(c)} four-row coupled fractal Cantor NPWs studied
in the present paper. \label{fig:PCWs}}
\end{figure}

Other light guiding structures are nanopillar waveguides (NPWs)
\cite{NWG1,NWG2}, one-dimensional (1D) periodic 
arrangements of dielectric rods (see Fig. \ref{fig:PCWs}), where the
guidance is realized by total internal reflection.  
They offer a variety of advantages over PCWs: 
The NPW geometry allows for more flexible arrangements than 
PCWs without substantially degrading the good transmission
\cite{NWGperf}, including bendings and arbitrary alterations of the longitudinal 
periodicity. 
It has been shown that microcavities with high $Q$-factors can be 
formed by point defects in an otherwise periodic NPW due to the effect
of multipole cancellation \cite{NWGJohnson}. Furthermore, the NPWs
allow for easy change of the pillar arrangement from periodic to quasiperiodic
or deterministically aperiodic (DA), as recently proposed by us \cite{NWGjosa}.
It has been shown that such DA NPWs support resonant modes which retain
a high $Q$-factor and have a spatial radiation loss profile in
favor of the coaxial direction (i.e. direction of periodicity or quasi-periodicity). 
These modes are intermediate between guided modes in
periodic NPWs and localized microcavity modes, as is characteristic
for non-periodic geometries.

Another distinct advantage of NPWs is the possibility to create
quasi-1D, i.e.
multiple-row coupled NPWs (Fig.~\ref{fig:PCWs}b) \cite{NWG1,NWG2}.
The modes of the waveguide are then split, forming a clear supermode
structure (see Fig.~\ref{fig:modes}). By varying the inter-row spacing
and lateral displacement, one can vary the coupling strength arbitrarily,
which is very difficult in PCWs. What is more, it was shown that the
supermodes can be easily excited selectively by placing the coherently
emitting dipoles according to the spatial mode configuration \cite{NWG1}.

In this paper we propose to use such a coupled NPW as a multimode 
laser cavity, when fabricated from a laser-active material. 
We investigate the possibility of \emph{switchable} lasing mode 
selection in such devices. In this setup a continuous pumping source 
drives the occupation inversion in the active medium and, hence, the 
laser action. The mode selection among several \emph{pre-existing modes} 
of an unaltered NPW device is then achieved by an additional optical 
pulse applied at the onset of the lasing action, 
using a spatial pulse intensity profile corresponding to the mode to 
be selected (optical injection seeding \cite{seeding}). 
This is an alternative to 
the known concept of tunable lasers where the resonator itself is 
modified, e.g. by thermal, electrooptical (using liquid crystals) or
micromechanical means \cite{tune1,tune2therm,tune3mech,tune4LC} 
to tune a given mode to have the desired frequency.
One of the advantages may be that modes selected by injection seeding 
can be better pre-engineered to     
meet the desired fundamental (e.g., cavity QED) or application purposes 
\cite{NWG1}.

The paper is organized as follows. Section~2 introduces the geometry
and resonant mode structure of the coupled NPWs under study as well as
the model of a four-level laser-active medium. 
In Section~3 we explain the
injection seeding mechanism and discuss the numerical simulation of
selective lasing. Section~4 summarizes the results.

\section{Structures and laser model}

\begin{figure}[b]
\begin{minipage}[b][1\totalheight]{0.5\columnwidth}%
\centerline{\includegraphics[scale=0.45]{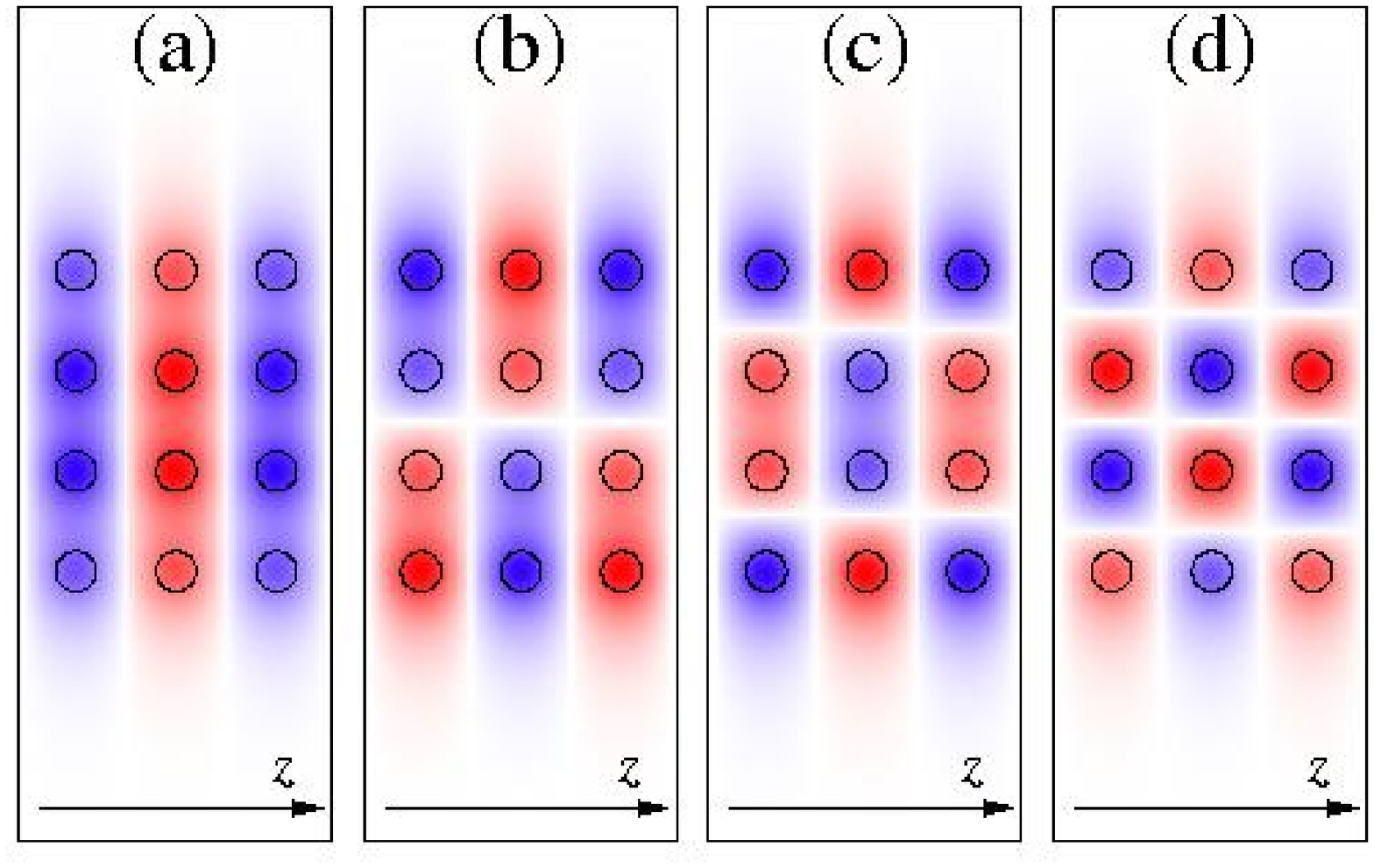}}

\caption{Lowest-order guided modes in a four-row infinite periodic NPW, calculated
with the plane wave expansion method. The shaded areas represent the
electric field strength with alternating signs.\label{fig:modes}}%
\end{minipage}%
\hfill{}%
\begin{minipage}[b][1\totalheight]{0.4\columnwidth}%
\noindent \begin{center}\includegraphics[bb=120bp 270bp 500bp 620bp,clip,width=0.88\columnwidth,angle=90]{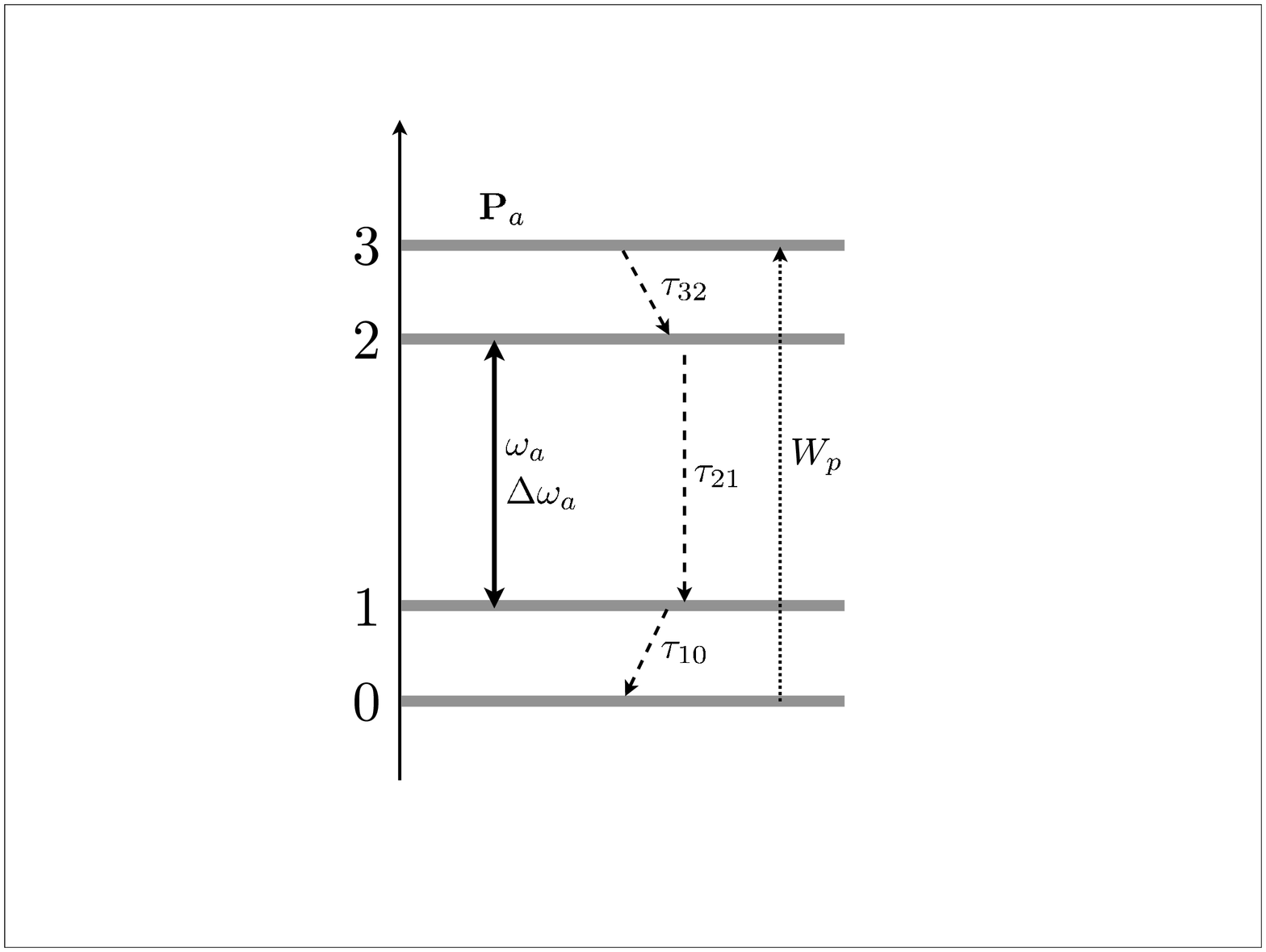}\par\end{center}

\caption{Schematic illustration of the four-level laser model with parameters
shown.\label{fig:levels}}%
\end{minipage}%

\end{figure}

We consider a coupled four-row periodic NPW (see Fig.~\ref{fig:PCWs}b),
as described in Ref.~\cite{NWG1}. The modes in question are shown in 
Fig.~\ref{fig:modes}. They are supermodes
formed by splitting of the first band edge resonance, resulting
from the finite extension of the structure in the $z$ direction. 
The $Q$-factor is the strongest for these modes, which means
that they dominate the NPW spectrum in the region of interest.
In addition, following the results in Ref.~\cite{NWGjosa} we consider
an analogous four-row NPW based on a fractal Cantor 
structure (Fig.~\ref{fig:PCWs}c),
which supports critically localized supermodes. One can notice that
the latter structure has two kinds of interpillar distances (short~$d_{S}$
and long~$d_{L}$) alternating in a non-periodic manner according
to a procedure similar to the triadic Cantor set formation. For details
on the construction procedure we refer the reader to Ref.~\cite{NWGjosa}.
Note that it is just one example of fractal structures; for more possibilities
the reader is referred to Ref.~\cite{EPL} and references therein.
In our numerical realization of NPWs we have used the following 
parameter values: The interpillar distances were taken to be
$d_{S}=0.5a$, $d_{L}=0.81a$ (the latter also used for periodic structures)
and the inter-row spacing $\Delta=1.5d_{L}=1.21a$.
The nanopillars with the radius $r=0.15a$ and dielectric constant
$\varepsilon=13$ were placed in air ($\varepsilon=1$). 
The model is fully scalable \cite{lasJoannop} 
with the length scale $a$ and simultaneously with the frequency 
scale $2\pi c/a$. For the explicit computations we used $a$=500~nm.

In order to model the gain in NPWs, we consider the
semiclassical four-level laser model with external (e.g., electrical)
pumping \cite{lasIEEE,lasRandom} (Fig.~\ref{fig:levels}). 
The lifetimes~$\tau_{32}$, $\tau_{21}$ 
and~$\tau_{10}$ characterize non-radiative transitions between corresponding
levels. Between the levels 2 and 1 there is also a radiative transition,
the lasing transition, with frequency ~$\omega_{a}$ and radiative 
linewidth $\Delta\omega_{a}\gg 1/\tau_{21}$. In order to achieve population inversion,
the lifetimes are taken 
so that $\tau_{32}\simeq\tau_{10}\ll\tau_{21}$.
The population dynamics is controlled by the standard rate equations,
\begin{eqnarray}
\frac{dN_{3}}{dt}&=&-\frac{N_{3}}{\tau_{32}}+W_{p}N_{0},\label{eq:rate3}\\
\frac{dN_{2}}{dt}&=&\frac{N_{3}}{\tau_{32}}-\frac{N_{2}}{\tau_{21}}+\frac{1}{\hbar\omega_{a}}\mathbf{E}\cdot\frac{\partial\mathbf{P}_{a}}{\partial t} \label{eq:rate2}\\ 
\frac{dN_{1}}{dt}&=&\frac{N_{2}}{\tau_{21}}-\frac{N_{1}}{\tau_{10}}-\frac{1}{\hbar\omega_{a}}\mathbf{E}\cdot\frac{\partial\mathbf{P}_{a}}{\partial t} \label{eq:rate1}\\ 
\frac{dN_{0}}{dt}&=&\frac{N_{1}}{\tau_{10}}-W_{p}N_{0}.\label{eq:rate0}
\end{eqnarray}
Here the external pumpimg is modeled by the phenomenological
pumping rate~$W_{p}$, which transfers electrons from the ground
level to the uppermost level \cite{lasIEEE,lasRandom}. The macroscopic
polarization $\mathbf{P}_{a}(\mathbf{r},t)$ 
of the medium is governed in each spatial point by the
equation of motion,
\begin{equation}
\frac{\partial^{2}\mathbf{P}_{a}(\mathbf{r},t)}{\partial t^{2}}+\Delta\omega_{a}\frac{\partial\mathbf{P}_{a}(\mathbf{r},t)}{\partial t}+\omega_{a}^{2}\mathbf{P}_{a}(\mathbf{r},t)=\sigma_{a}\mathbf{E}(\mathbf{r},t)\Delta N(\mathbf{r},t)\label{eq:polarization}\end{equation}
where $\Delta N=N_{1}-N_{2}$ is the population inversion, and the
coefficient~$\sigma_{a}$ describes the coupling of $\Delta N$
to the electric field $\mathbf{E}$ produced by stimulated photon emission. 
It is known to be $\sigma_{a}=(6\pi\epsilon_{0}c^{3})/(\tau_{21}\omega_{a}^{2})$
\cite{lasIEEE,lasPauli}. The electromagnetic fields are governed
by the usual Maxwell equations,
\begin{equation}
\nabla\times\mathbf{E}=-\mu_{0}\mu\frac{\partial\mathbf{H}}{\partial t},\quad\nabla\times\mathbf{H}=\varepsilon_{0}\varepsilon\frac{\partial\mathbf{E}}{\partial t}+\frac{\partial\mathbf{P}_{a}}{\partial t}+\mathbf{j}(t).\label{eq:maxwell}\end{equation}
We solve the system of equations (\ref{eq:rate3}-\ref{eq:maxwell}) 
by the finite-difference time-domain method \cite{AVLCharact}
with auxiliary differential equations (FDTD-ADE), e.g., as described
in \cite{lasIEEE,lasRandom}. 
The external current density~$\mathbf{j}(t)$ is introduced in 
Eq.~(\ref{eq:maxwell}) as a technical tool within the FDTD method
to excite the electromagnetic fields in the system 
(see Section 3). For the dielectric medium at hand $\mathbf{j}(t)$ 
is assumed to be Gaussian pulses of oscillating point dipole sources
throughout this paper,  
\begin{equation}
\mathbf{j}(t)\sim\exp\left[\frac{(t-t_{0})^{2}}{\sigma_{t}^{2}}\right]\sin\omega t.\label{eq:pulse_gauss}\end{equation}
For the explicit FDTD solutions the nanopilars were assumed to extend
infinitely in the third dimension ($y$-axis), and
TM polarization was considered, so that the electric field has only a
$y$-component, $E(\mathbf{r},t)\equiv E_{y}(x,z,t)$.
To model an open system, perfectly matched layer (PML) boundary conditions
were used \cite{AVLCharact}. The computational domain of 
size $7a\times22a$ was discretized with 16 mesh points within the unit
length $a$. The time step is connected to the spatial mesh size to
assure stability and was taken to be $dt=1/16 (a/\sqrt{3}c)$, i.e.
$dt=0.60182\times10^{-16}\textrm{s}$ for $a$=500~nm.
The four-level model parameters were chosen as 
$\tau_{31}=\tau_{10}=1.0\times10^{-13}\textrm{s}$,
$\tau_{21}=3.0\times10^{-10}\textrm{s}$, 
and the total numer of laser-active atoms was taken to be
$N_{\textrm{total}}\equiv\sum _{i=0\dots 3} N_i = 10^{24}$ 
per computation cell, as used in \cite{lasPauli}.

\begin{figure}
\includegraphics[
clip,width=0.2\columnwidth]{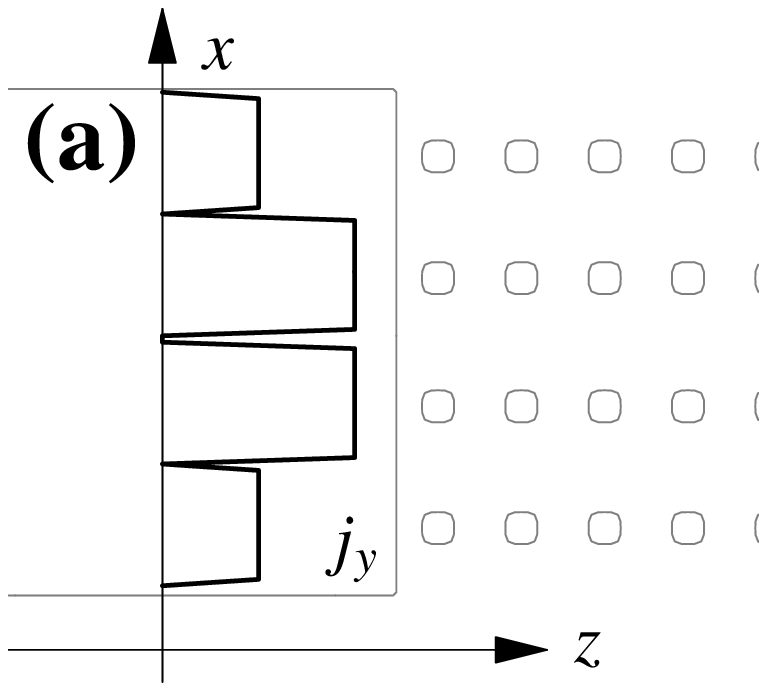}\hfill{}
\includegraphics[
clip,width=0.2\columnwidth]{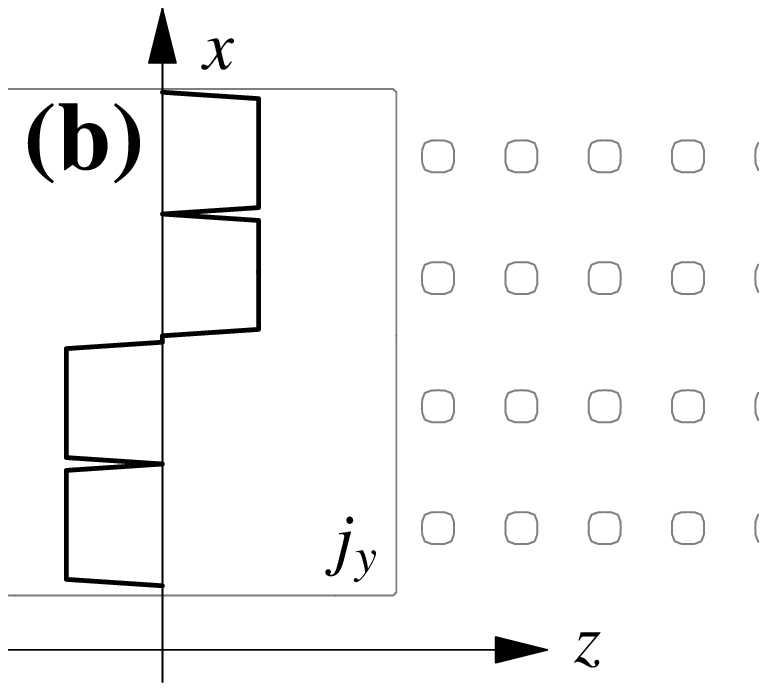}\hfill{}
\includegraphics[
clip,width=0.2\columnwidth]{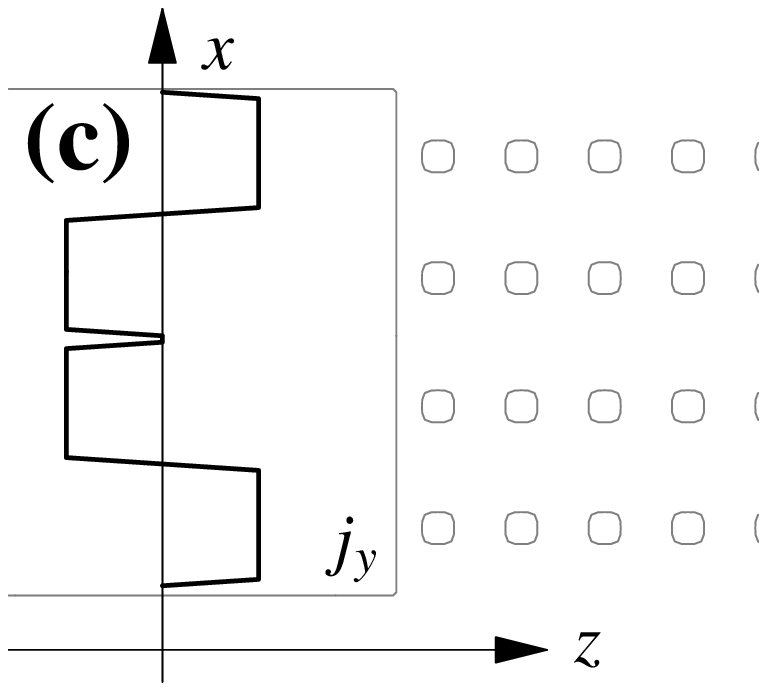}\hfill{}
\includegraphics[
clip,width=0.2\columnwidth]{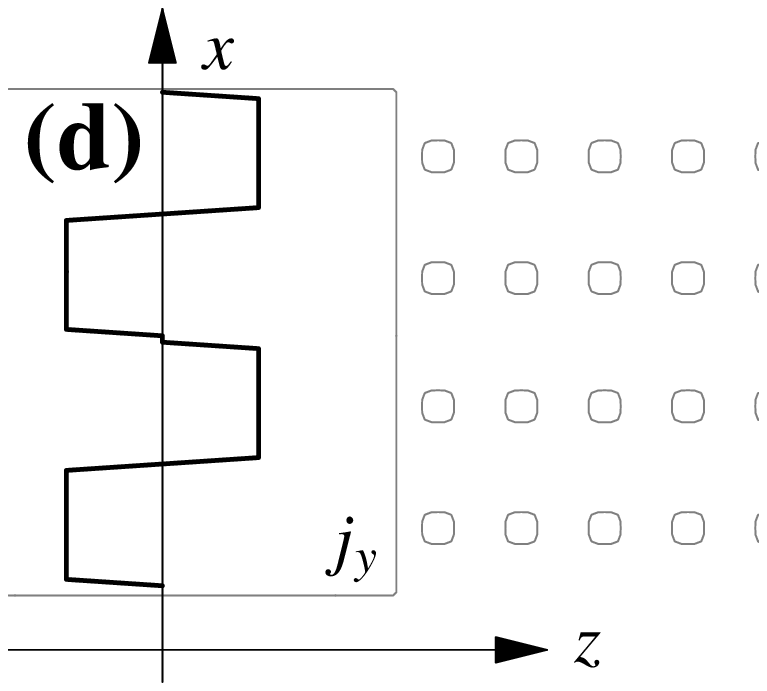}

\caption{The fragment of the four-row waveguide with the coaxial terminal along
with the schematic illustration of the transverse envelope of the
seeding signal $j_{y}(x)$ as excited in the terminal. \label{fig:seeding}}
\end{figure}

\section{Injection seeding and selective lasing}

As a microlaser device with injection seeding we consider a
periodic or Cantor NPW (as in Fig.~\ref{fig:PCWs})
coupled to a terminal through which the seeding signal
is injected. 
The terminal is a slab wavewaveguide which is 
coaxially aligned with the NPW and extends out to the PML boundary 
of our system. 

The externally pumped laser-active medium 
[Eqs.~(\ref{eq:rate3})-(\ref{eq:rate0})]
is placed in the pillars of all four rows of
the central 10-pillar group (Cantor) or the 7 centermost  
pillars (periodic) of the NPW.
This is done to maximize coupling between the active medium and the
main localization region of the lasing modes; in the periodic case,
it also helps to reduce the influence of second-order Fabry-P\'erot
modes.

The seeding signal is excited by four actuators, i.e., linear groups of 
point dipole sources, modeled by Eq.~(\ref{eq:pulse_gauss}) 
placed at the outer end of the terminal 
(next to the PML boundary) and aligned perpendicularly to the waveguide.
Each of the actuators emits
a single, short Gaussian pulse \eqref{eq:pulse_gauss}
with carrier frequency~$\omega$ at or near~$\omega_{a}$ and with a
half-width duration $\sigma_{t}=10^{4}dt$. 
The relative phases and amplitudes of the
actuators were taken to produce four different electric field 
profiles as shown in Fig.~\ref{fig:seeding} so as to excite the four
different-order transverse modes in the slab waveguide and in the NPW as
shown in Fig.~\ref{fig:modes}. Since the mode selection is based
on the {\it coherent} light emission from the externally pumped medium 
stimulated by the seeding field, the mode selection process is
phase sensitive \cite{Siegman,seeding},  
that is, e.g. the modes of Fig.~\ref{fig:modes}
can be distinguished, because they differ in the phase distribution
but have essentially the same spatial intensity distribution. 
However, the selectivity may degrade, if the seeding signal induced
by the actuators is not a waveguide eigenmode, so that several modes
are excited in the waveguide while the signal propagates from the
terminal to the laser-active region. 

\begin{figure}

\hfill{}\includegraphics[width=0.5\columnwidth]{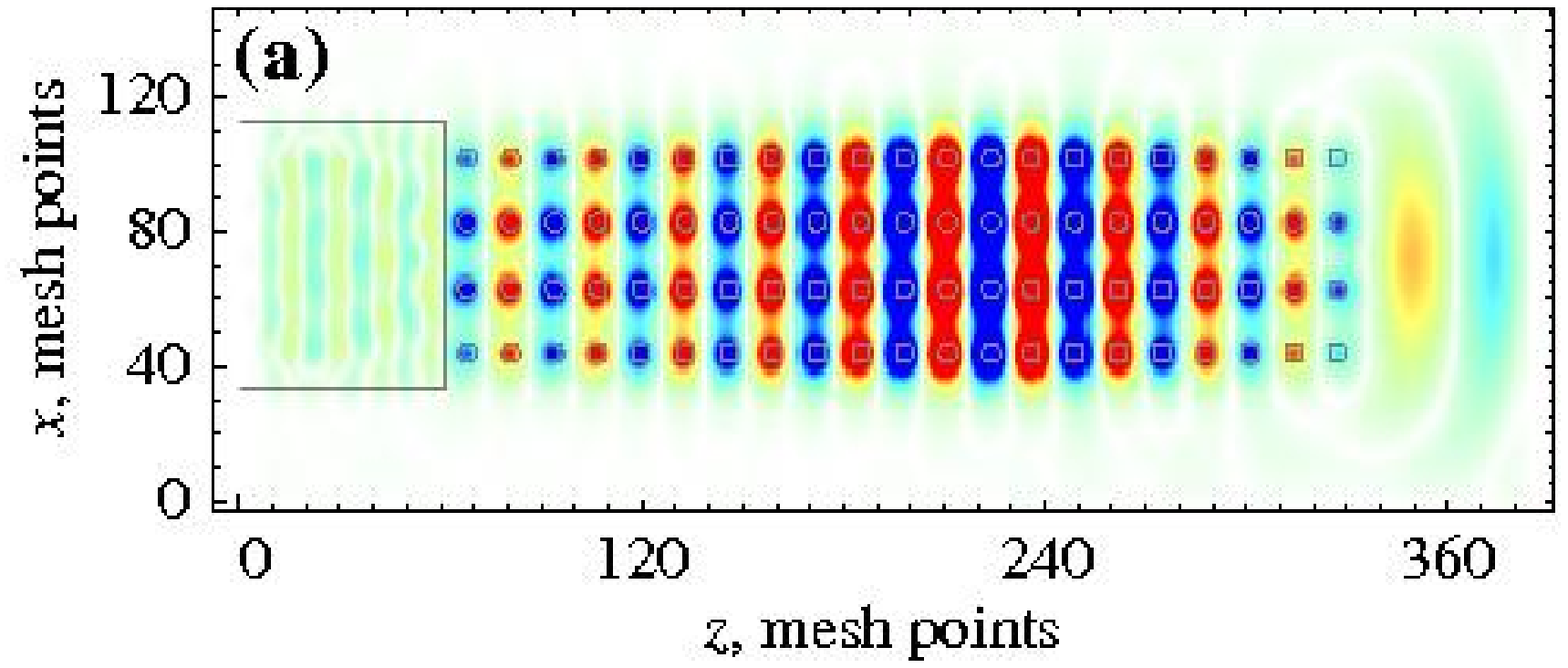}\hfill{}\includegraphics[width=0.5\columnwidth]{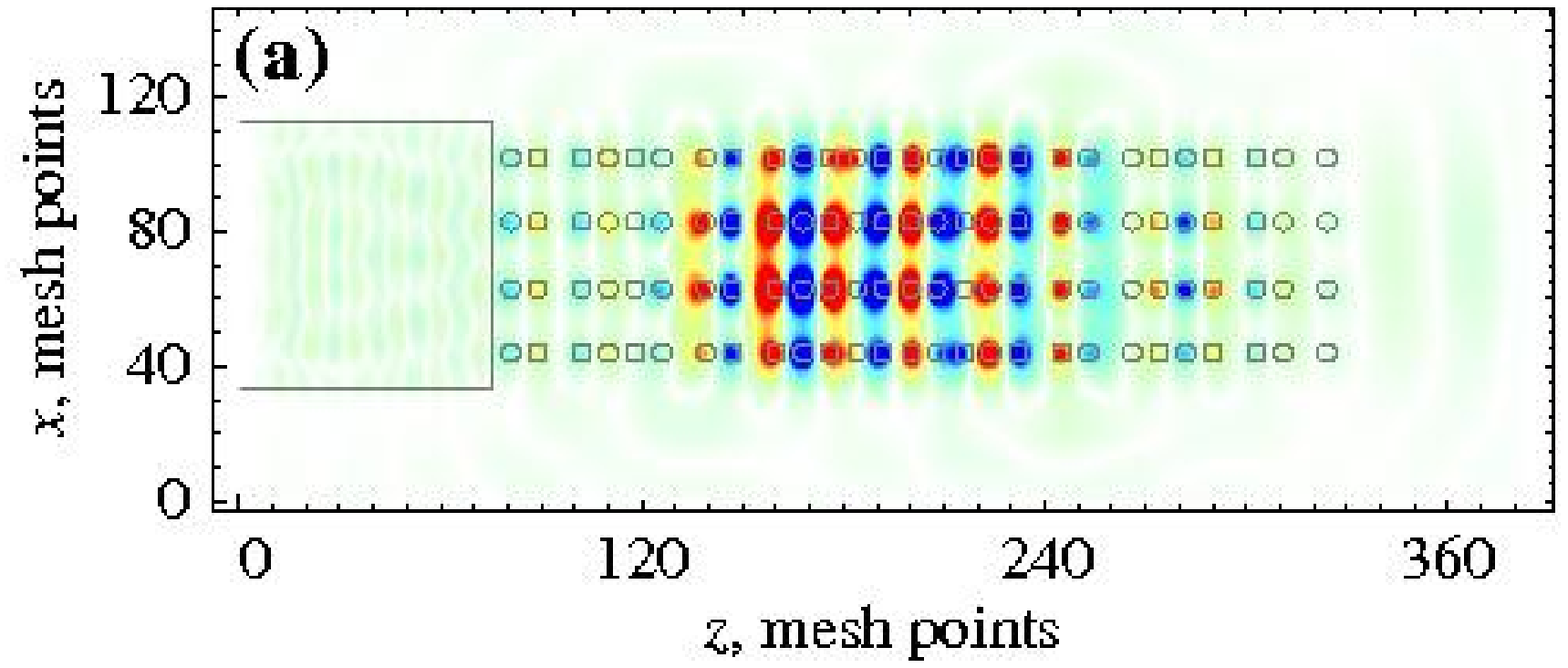}\hfill{}

\hfill{}\includegraphics[width=0.5\columnwidth]{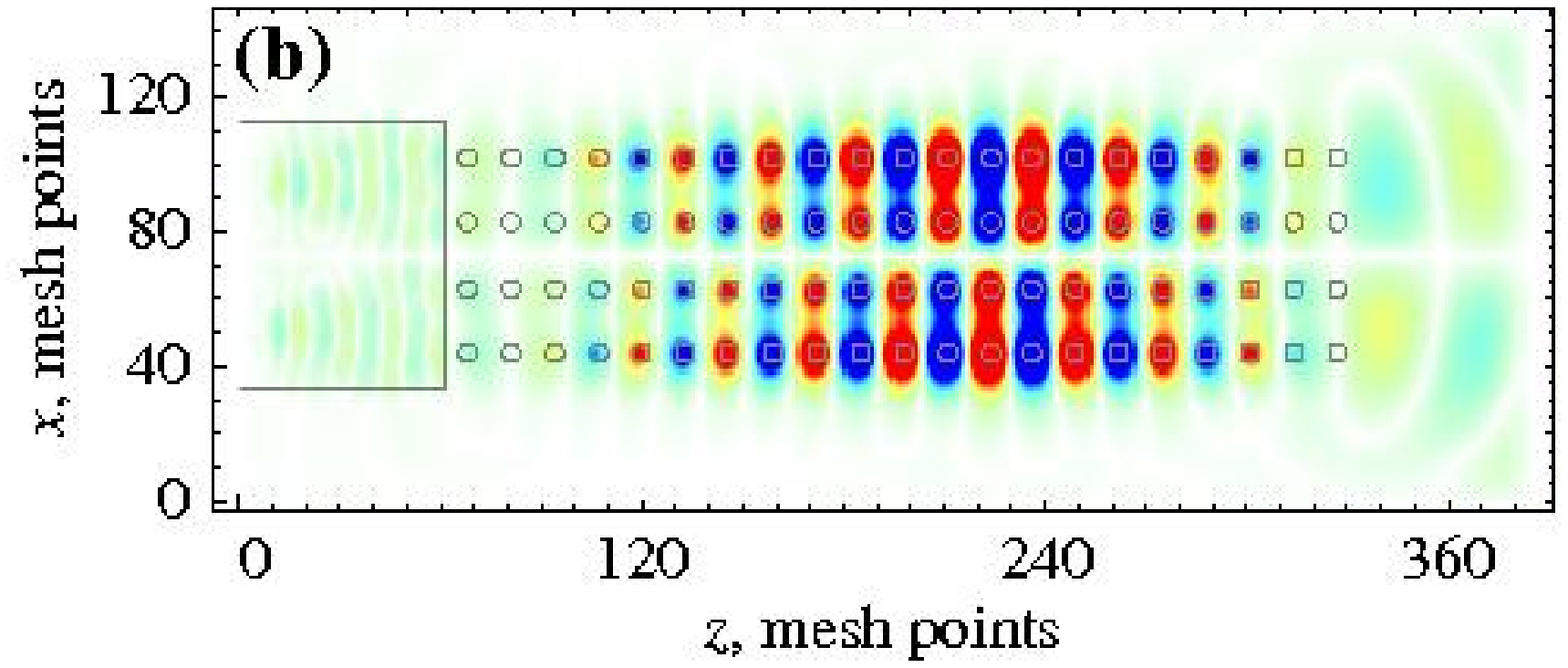}\hfill{}\includegraphics[width=0.5\columnwidth]{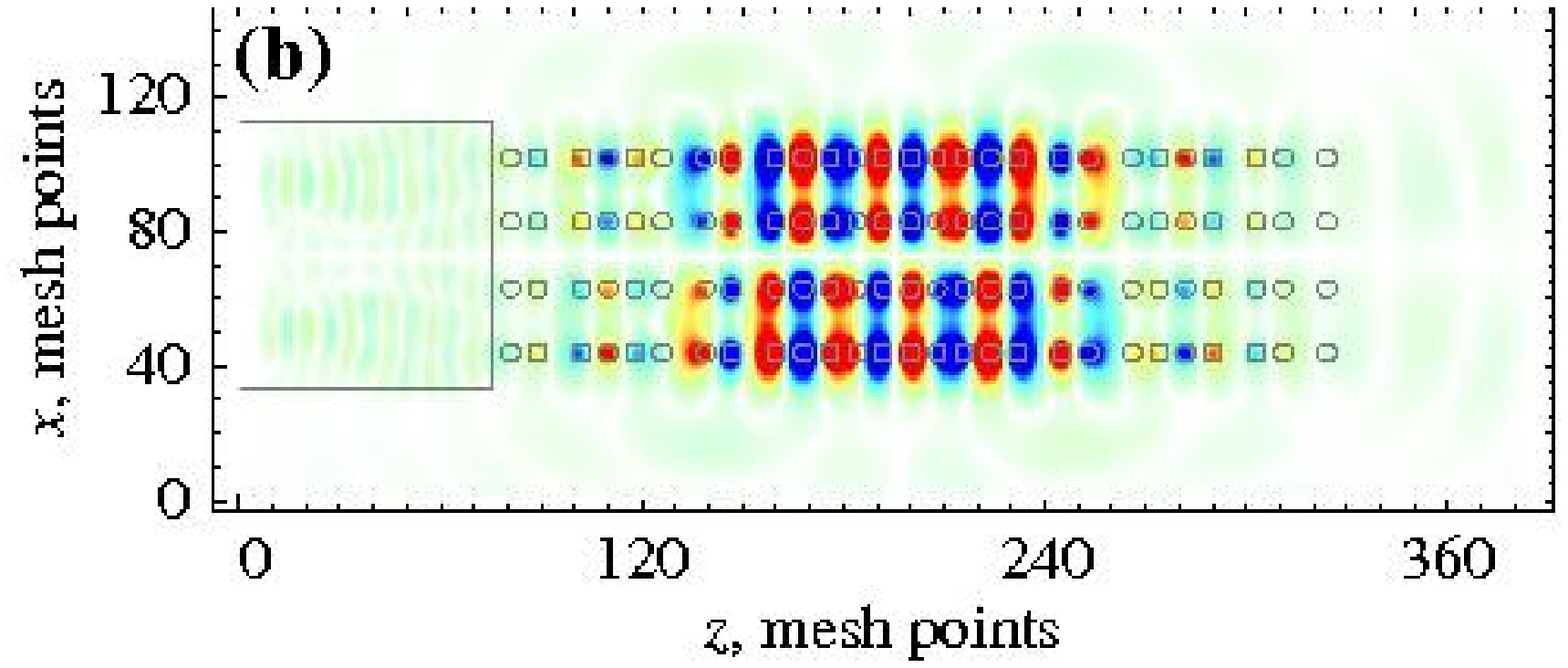}\hfill{}

\hfill{}\includegraphics[width=0.5\columnwidth]{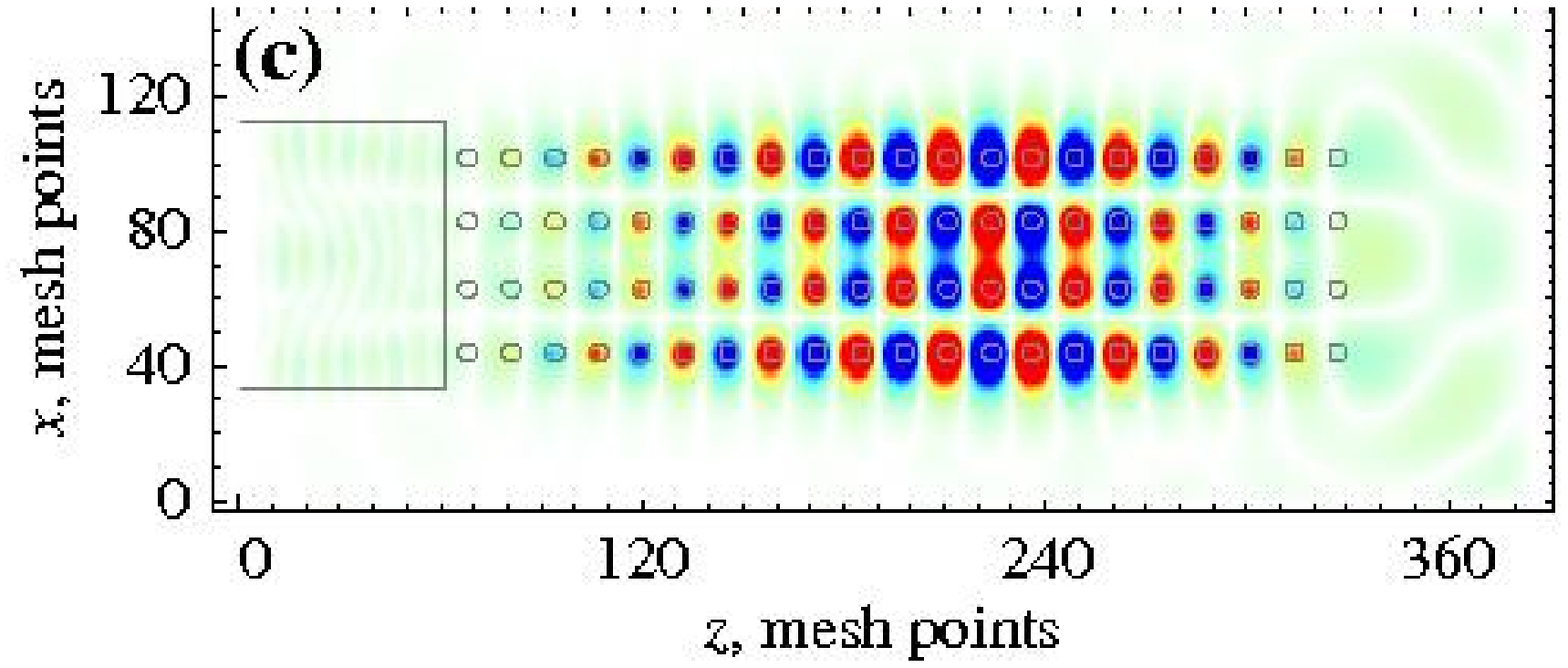}\hfill{}\includegraphics[width=0.5\columnwidth]{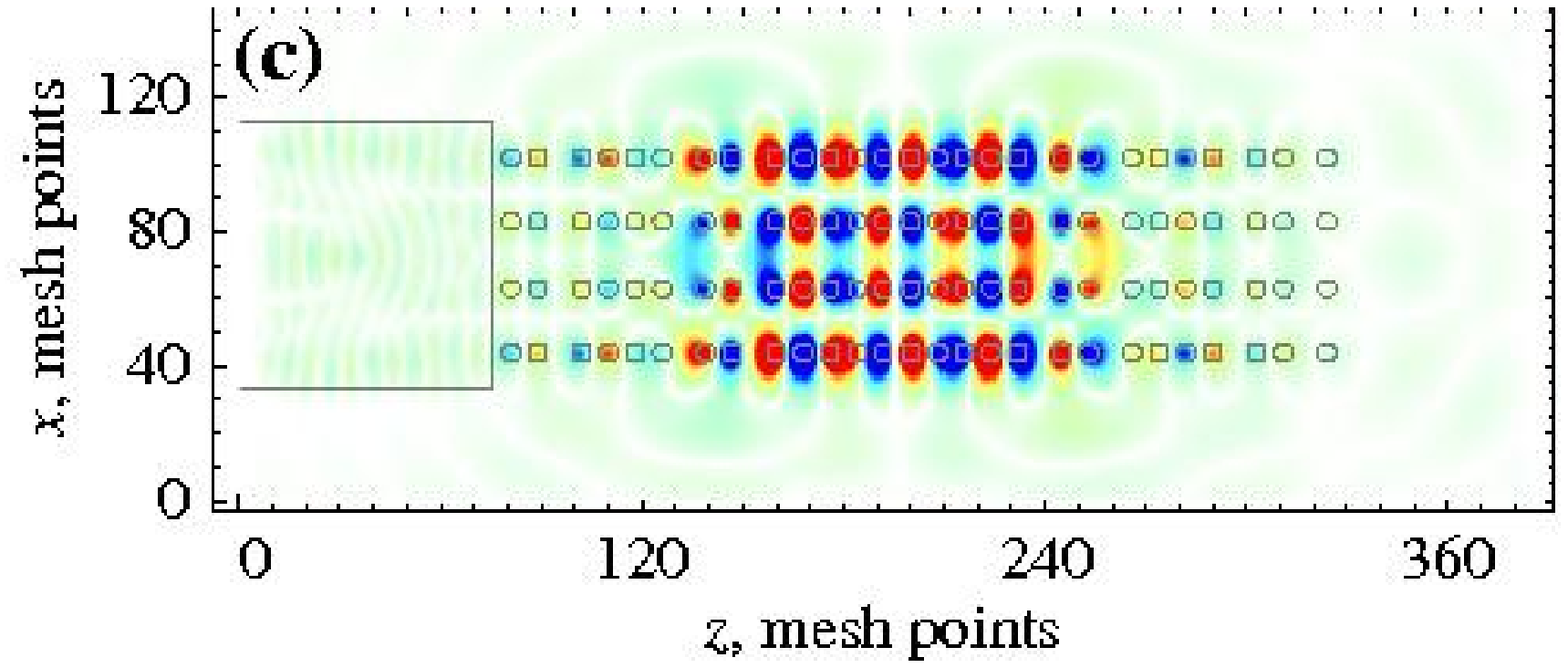}\hfill{}

\hfill{}\includegraphics[width=0.5\columnwidth]{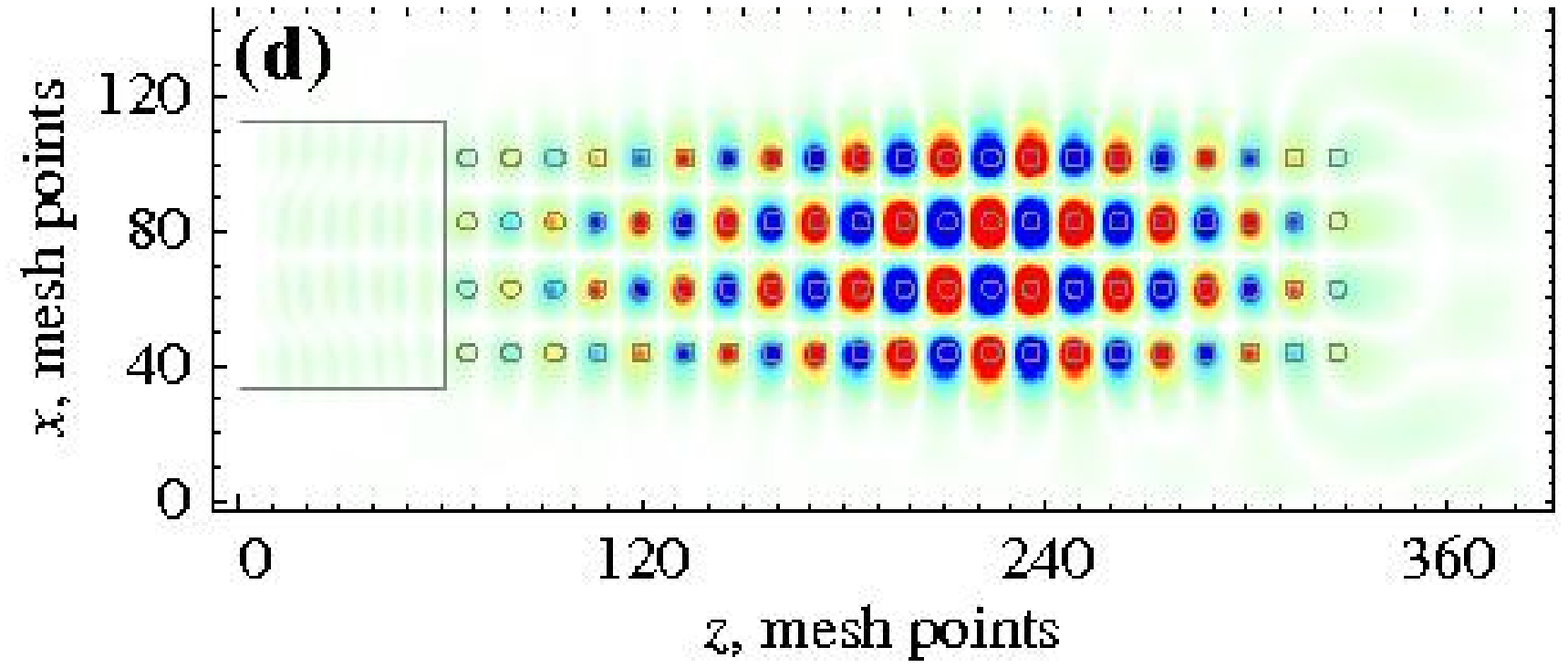}\hfill{}\includegraphics[width=0.5\columnwidth]{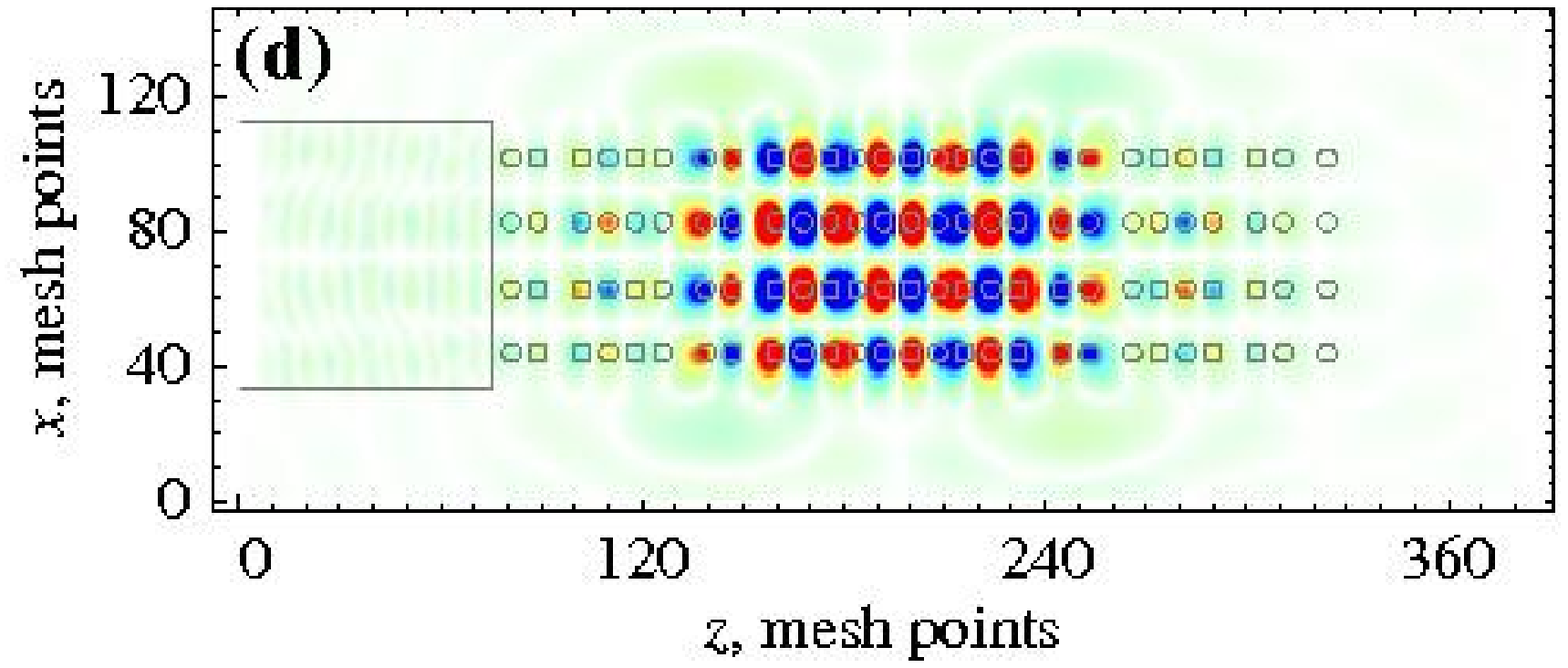}\hfill{}

\caption{The electric field distribtion in four different modes 
for externally pumped, injection-seeded
periodic (left panel) and fractal Cantor (right panel) four-row NPWs.
The central lasing frequency is $\omega_{a}=0.3185)$ (periodic) 
and $0.364\,(2\pi c/a)$ (Canotor), respectively.
The panels (a)--(d) correspond to the seeding signals
(a)--(d) in Fig.~\ref{fig:seeding}. Note the
similarity to corresponding infinite NPW modes 
in Fig.~\ref{fig:modes}.\label{fig:field}}
\end{figure}

The mode selectivity is influenced by three major factors,
(1) the precise time $t_0$ at which the Gaussian seeding signal is
applied relative to the switch-on time of the pumping power;
(2) random noise, generated, e.g., by spontaneous emission during
the selection process, and
(3) the $Q$-factors as well as the coupling strengths of the different
modes to the lasing transition of the active material.
Clearly, the seeding signal must be applied during the transient
time interval, i.e. after the pumping power has been switched on,
but before steady-state lasing has been reached, since otherwise the
lasing would always be dominated by the one mode coupled most strongly 
to the lasing transition, independent of the seeding signal. 
This temporal dynamics and the influence of random noise 
will be analyzed in a forthcoming publication. 

\begin{figure}
\includegraphics[clip,width=0.5\textwidth,keepaspectratio,angle=-90]{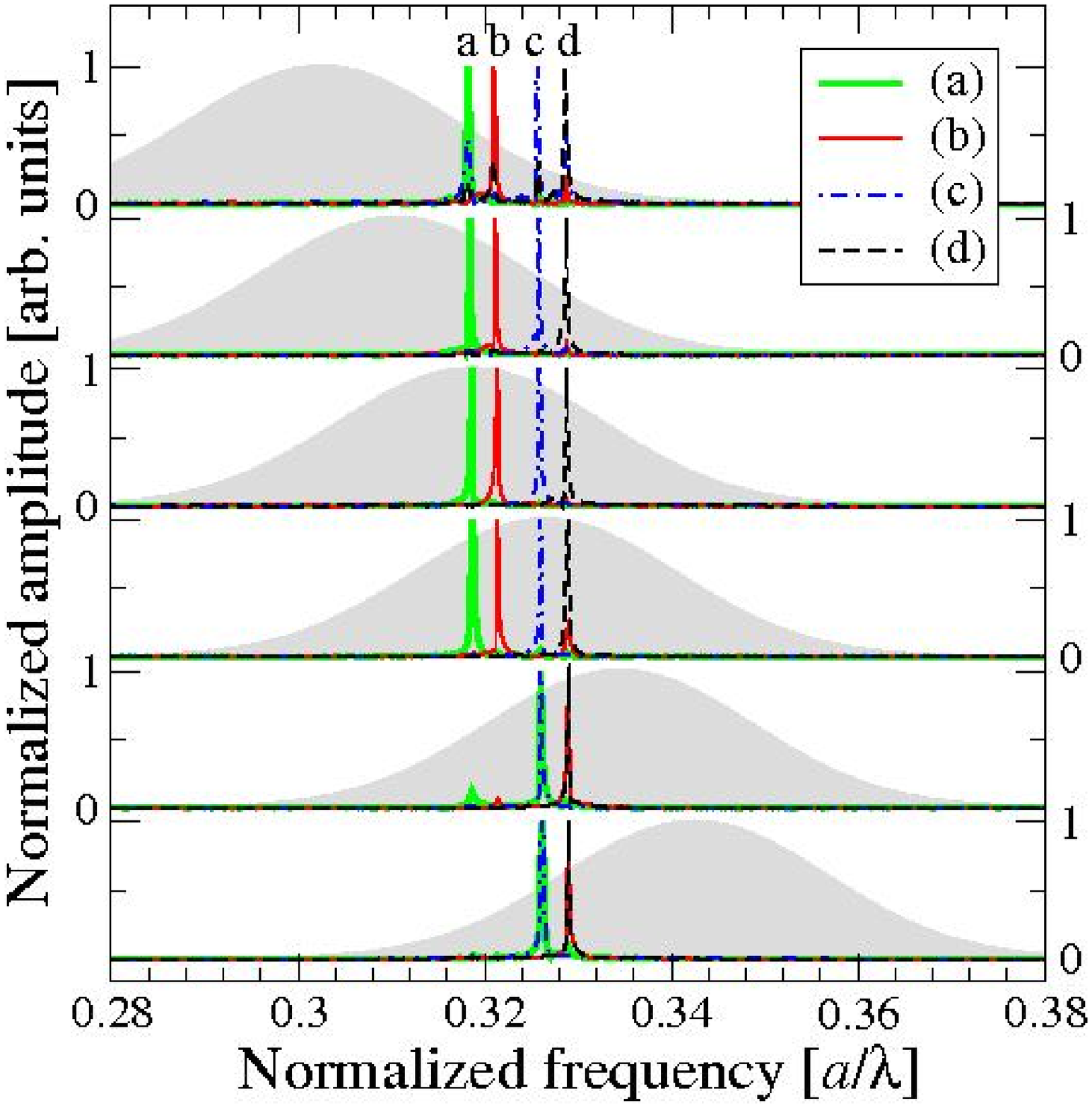}\hfill{}\includegraphics[clip,width=0.5\textwidth,keepaspectratio,angle=-90]{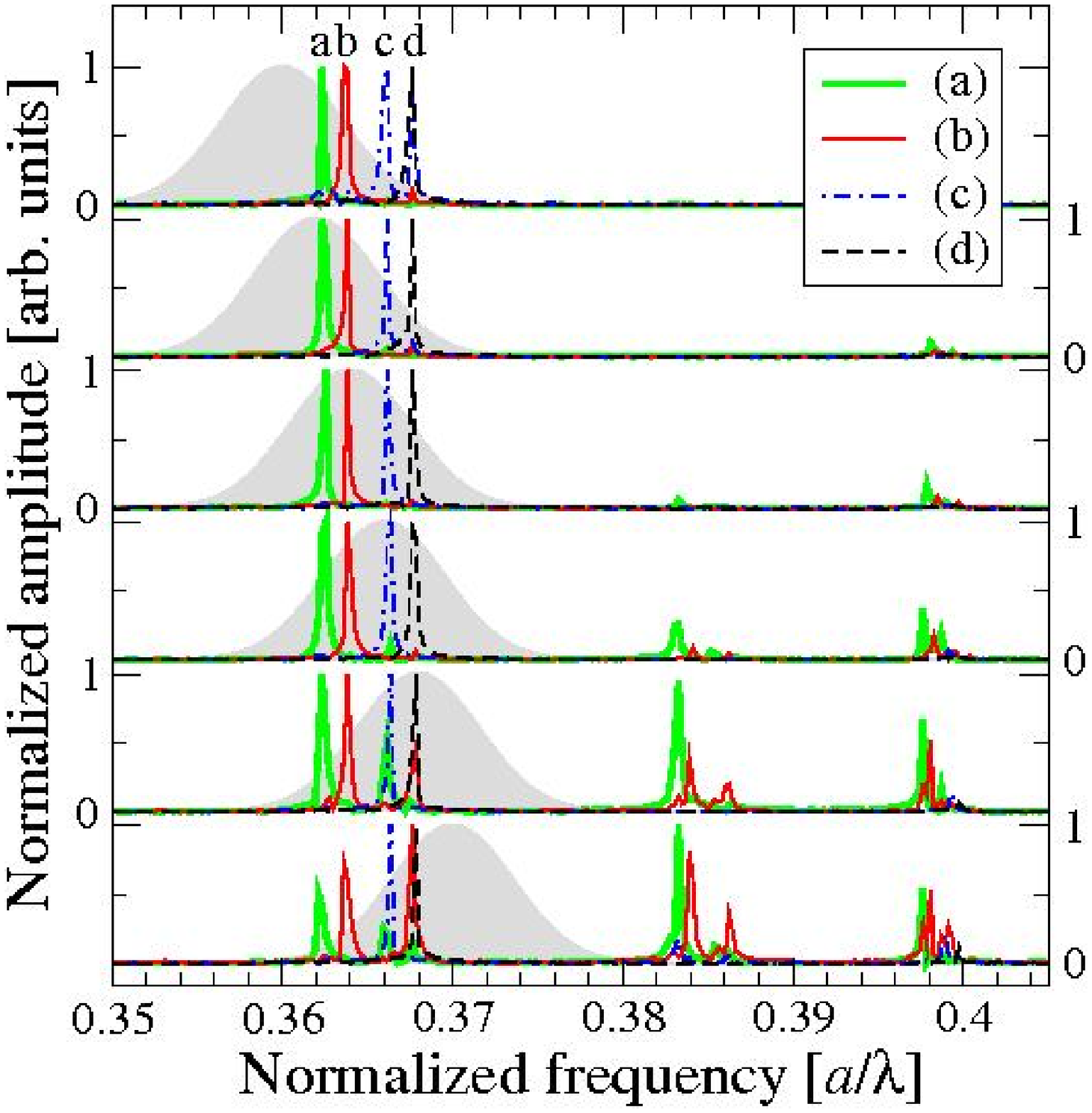}

\caption{The amplitude spectra (each one normalized to its own maximum) of
the electric field in periodic (left panels) and 
Cantor (right panels),
externally pumped, injection-seeded four-row NPWs. 
The lines labeled \textbf{(a)}
through \textbf{(d)} correspond to the seeding signals (a)--(d) 
shown in Fig.~\ref{fig:seeding}. The shaded areas represent
the laser amplification line, with its central frequency~$\omega_{a}$
varying in equidistant steps from 0.3025 to 0.3425 $(2\pi c/a)$ for periodic
structures and from 0.36 to 0.37 $(2\pi c/a)$ for Cantor structures
(top to bottom panels).
The pumping rate equals $W_{p}=1.0\times10^{13}\,\textnormal{s}^{-1}$.
\label{fig:seed-omega}}
\end{figure}

Here we focus on the mode selection in depencence of the $Q$-factors and
coupling strengths of the modes. In Fig.~\ref{fig:field} we show the 
spatial electric field distribution in the four-row NPW laser device 
described above at an instant of time long after the seeding signal has
decayed and after the steady state has been reached. Here the frequency of the
lasing transition $\omega_a$ was chosen so that the laser line overlaps with 
the mode eigenfrequencies. The symmetry of the
selected lasing modes in Fig.~\ref{fig:field} corresponds to that of the 
seeding signal shown in Fig.~\ref{fig:modes}. It is seen that the lasing modes
are the better localized inside the NPW structure (and thus their
$Q$-factor is the higher) the higher their 
transverse order and, hence, the higher their frequency. Therefore, the 
higher-order modes are expected to be favored for lasing. This effect
can be compensated for by different couplings of those modes to the lasing
transition, as has been
further analyzed in Fig.~\ref{fig:seed-omega}. It shows the lasing 
spectra (sharp coulored lines) in the steady state long after the seeding 
signal has decayed for various laser transition frequencies $\omega_a$.  
The broad shaded areas depict the laser line of width $\Delta \omega_a$
centered at $\omega_a$. When the mode 
eigenfrequencies lie slightly above $\omega_a$ (but still overlapping with 
the laser line), any of the four NPW modes can be selected by the 
appropriate seeding signal with the same symmetry, as seen  
in Fig.~\ref{fig:seed-omega}, upper four panels. We attribute this to
the fact that in this case the low-frequency NPW modes 
have a stronger spectral overlap with the laser line than the  
high-frequency modes, thus compensating for the larger losses in the
low-frequency modes. When the NPW mode frequencies are below the
laser line (Fig.~\ref{fig:seed-omega}, lower two panels), the opposite 
effect occurs. In this case, for periodic NPWs (left panels)
the two low-order modes cannot be 
selected by injection seeding, because of too low $Q$-factor
and too weak coupling to the laser line. For Cantor NPWs 
the situation is somewhat better (right panels), presumably because
in the Cantor structures the modes are better localized inside the
NPW and have considerably higher $Q$-factors, which are less strongly
dependent on the order of the modes, as seen 
in Fig.~\ref{fig:field} and analyzed in Ref.~\cite{NWGjosa}.  
It is also seen in Fig.~\ref{fig:seed-omega} that in Cantor NPWs other modes
than the four fundamental ones can participate in the lasing. Further
analysis is needed to find out if this is due to their different
(critical) spatial localization properties \cite{NWGjosa}, 
which might prevent them from being subject to mode competition. 

\section{Conclusions and outlook}

We have numerically investigated the possibility of selective lasing
into each of the modes formed in coupled resonators \cite{NWG1}.
Coupled nanopillar waveguides of both periodic and deterministically
aperiodic fractal geometries have been used as a model. We have found
that injection seeding can be used 
in order to achieve selective lasing with an externally pumped active
medium. The seeding signal must be applied during the relatively short time
period of the onset of lasing. The dependence of the mode selection on various
parameters such as the lasing frequency has been investigated. We
have shown that in fractal structures the modes are better localized
than in the periodic case and yet show a good coupling to a coaxial
terminal (compare graphs in Fig.~\ref{fig:field}) \cite{NWGjosa}. 
As a consequence,
the lasing performance in such structures is expected to be greater,
as conjectured by us earlier \cite{NWGjosa}. 
Note that the nanopillar-based
structures in the wavelength range considered ($1\div1.5\,\mu\textnormal{m}$)
are within the state-of-the-art fabrication possibilities, both in
sandwich-like \cite{JapPillar} and in membrane-like \cite{tune3mech}
geometry.

Based on the results obtained, we propose a new concept 
of \emph{switchable}
(rather than tunable) lasing in microstructures, when instead of changing
the parameters of a single-mode cavity an inherently multiple-mode
resonator is used, and one of the modes is deliberately selected
for lasing. The modes can be pre-engineered to have the desired properties,
and 1D deterministically aperiodic structures considered here
offer broad possibilities for such engineering. However, the results
presented here only form the first step towards such a concept. 
More detailed results and analyses will be presented in
forthcoming publications. 

Furthermore, in order to consistently develop the proposed concept
of mode selection or switching, a theoretical description of lasing
in a multimode resonator with arbitrary mode characteristics is required.
Special attention needs to be given to interaction of the modes with
different localization character, as well as modes having spatial
and/or spectral overlap. Such a description has not yet been made,
and is a subject for future studies. 

\begin{acknowledgement}
This work was supported by the Deutsche Forschungsgemeinschaft (SPP
1113 and FOR 557).
\end{acknowledgement}


\begin{thebibliography}{20}
\bibitem{Joannopoulos}J.~D.~Joannopouos, \emph{}R.~D.~Meade,
and J.~N.~Winn, \emph{Photonic crystals: Molding the Flow of Light},
Princeton University Press (Princeton, 1995).

\bibitem{Sakoda}K.~Sakoda, \emph{Optical Properties of Photonic
Crystals}, Springer (Berlin, 2001).

\bibitem{Loutrioz}J.-M.~Loutrioz et al., \emph{Photonic Crystals:
Towards Nanoscale Photonic Devices}, Springer  (Berlin, 2005).

\bibitem{JohnsonGuide1}S.~G.~Johnson et al., \emph{Phys.~Rev.~B}
\textbf{60}, 5751 (1999).

\bibitem{JohnsonGuide2}S.~G.~Johnson, P.~R.~Villeneuve, S.~Fan and
J.~D.~Joannoupoulos, \emph{Phys.~Rev.~B} \textbf{62}, 8212
(2000).

\bibitem{JapPillar}M.~Tokushima, H.~Yamada and 
Y.~Arakawa, \emph{Appl.~Phys.~Lett.}
\textbf{8}4, 4298 (2004).

\bibitem{mcNoda1}Y. Akahane, T. Asano, B.-S. Song and S. Noda, \emph{Nature}
\textbf{425}, 944 (2003).

\bibitem{mcLasing}O. Painter et al., \emph{Science} \textbf{284},
1819 (2004).

\bibitem{mcNoda2}S. Noda, A. Chutinan and M. Imada, \emph{Nature} \textbf{407},
608 (2000).

\bibitem{NWG1}D. N. Chigrin, A. V. Lavrinenko and C. M. Sotomayor-Torres,
\emph{Opt. Express} \textbf{12}, 617 (2004).

\bibitem{NWG2}D. N. Chigrin. A. V. Lavrinenko and C. M. Sotomayor-Torres,
\emph{Opt. Quantum Electron.} \textbf{37}, 331 (2005).

\bibitem{NWGperf}P.-G. Luan and K.-D. Chang, \emph{Opt. Express} \textbf{14},
3263 (2006).

\bibitem{NWGJohnson}S.~G.~Johnson, S.~Fan, A.~Mekis and J.~D.~Joannopoulos,
\emph{Appl.~Phys.~Lett.} \textbf{78}, 3388 (2001).

\bibitem{NWGjosa}S. V. Zhukovsky, D. N. Chigrin and J. Kroha, \emph{J.~Opt.
Soc.~Am.~B} \textbf{23}, 2265 (2006).

\bibitem{seeding}W.~Lee and W.~R.~Lempert, \emph{Appl.~Opt.} \textbf{42},
4320 (2003).

\bibitem{tune1}A.~Figotin, Y.~A.~Godin and I.~Vitebsky, \emph{Phys.~Rev.~B}
\textbf{57}, 2841 (1998).

\bibitem{tune2therm}K.~Yoshino et al., \emph{Appl.~Phys.~Lett.}
\textbf{75}, 932 (1999).

\bibitem{tune3mech}W.~Park and J.-B.~Lee, \emph{Appl. Phys. Lett.}
\textbf{85}, 4845 (2004).

\bibitem{tune4LC}E.~P.~Kosmidou, E.~E.~Kriezis and T.~D.~Tsiboukis,
\emph{IEEE J.~Quant. Electron.} \textbf{41}, 657 (2005).

\bibitem{EPL}S.~V.~Zhukovsky, A.~V.~Lavrinenko and S.~V.~Gaponenko,
\emph{Europhys. Lett.} \textbf{66}, 455 (2004).

\bibitem{lasJoannop}P.~Bermel, E.~Lidorikis, Y.~Fink and J.~D.~Joannopoulos,
\emph{Phys.~Rev.~E} \textbf{73}, 165125 (2006).

\bibitem{lasIEEE}A.~S.~Nagra and R.~A.~York, \emph{IEEE Trans. Ant.~Propag.}
\textbf{46}, 334 (1998).

\bibitem{lasRandom}X.~Jiang and C.~M.~Soukoulis, \emph{Phys.~Rev.~Lett.}
\textbf{85}, 70 (2000).

\bibitem{lasPauli}S.-H.~Chang and A.~Taflove, \emph{Opt.~Express}
\textbf{12}, 3827 (2004).

\bibitem{AVLCharact}A.~V.~Lavrinenko et al., \emph{Opt.~Express}
\textbf{12}, 234 (2004).

\bibitem{Siegman}A.~Siegman, \emph{Lasers}, University Science Books,
(Mill Valley, CA, 1986).

\end{thebibliography}
\end{document}